\def\dert#1#2{\frac{{{d}}{#1}}{{{d}}{#2}}}
\def\eqi{\begin{equation}}
\def\eqf{\end{equation}}
\def\rfr#1{eq. (\ref{#1})}

\def\virg#1{``#1''}
\def\rp#1#2{{#1\over#2}}
\def\lb#1{\label{#1}}
\def\beq{\begin{equation}}
\def\eeq{\end{equation}}
\documentclass{ws-ijmpd}
\usepackage{latexsym,wasysym}
\usepackage{graphicx,epsfig}
\RequirePackage{color}

\begin{document}

\markboth{L. Iorio \& M. L. Ruggiero}
{Phenomenological Constraints on the Kehagias-Sfetsos Solution}

%
%

\title{ HO\v{R}AVA-LIFSHITZ GRAVITY:   TIGHTER COSTRAINTS FOR THE KEHAGIAS-SFETSOS SOLUTION  FROM NEW SOLAR SYSTEM DATA}

\author{L. IORIO}

\address{Ministero dell'Istruzione, dell'Universit\`{a} e della Ricerca (M.I.U.R.). Permanent address: Viale Unit\`{a} di Italia 68\\
Bari, (BA) 70125,
Italy\\
lorenzo.iorio@libero.it}

\author{M. L. RUGGIERO}

\address{
Dipartimento di Fisica, Politecnico di Torino, Corso Duca degli Abruzzi 24\\  Torino, (TO) 10129, Italy}
\address{INFN, Sezione di Torino, Via Pietro Giuria 1 \\ Torino, (TO) 10125, Italy \\ matteo.ruggiero@polito.it}

\maketitle

\begin{history}
\received{13 December 2010}
\revised{2 March 2011}
\accepted{4 March 2011}
\comby{Ruth Gregory, Ph. D}
\end{history}

\begin{abstract}
We analytically work out the perturbation $\Delta\rho$ induced by the Kehagias-Sfetsos (KS) space-time solution of the Ho\v{r}ava-Lifshitz (HL) modified gravity at long distances on the two-body range $\rho$ for a pair of test particles A and B orbiting the same  mass $M$. We apply our results to the most recently obtained range-residuals $\delta\rho$ for some planets of the solar system (Mercury, Mars, Saturn) ranged from the Earth to effectively constrain the dimensionsless KS parameter $\psi_0$ for the Sun.
We obtain $\psi^{\odot}_0\geq 7.2\times 10^{-10}\ ({\rm Mercury}),\ \psi^{\odot}_0\geq 9\times 10^{-12}\ ({\rm Mars}),\ \psi^{\odot}_0\geq 1.7\times 10^{-12}\ ({\rm Saturn})$. Such lower bounds are tighter than other ones   existing in literature by several orders of magnitude. We also preliminarily obtain $\psi_0^{\bullet}\geq 8\times 10^{-10}$ for the system constituted by the S2 star orbiting the Supermassive Black Hole (SBH) in the center of the Galaxy.
\end{abstract}



\section{Introduction}

Since their publication, the Ho\v{r}ava's seminal papers\cite{horava1,horava2} on a four-dimensional theory of gravity which is power-counting renormalizable and, hence, can be considered a candidate for the short-range (UV) completion of the General Theory of Relativity (GTR), have raised a lot of interest and  stimulated an intense research effort on many topics of the theory and its modifications (a brief review  can be found in the recent paper by Sotiriou\cite{sotiriou}).  Ho\v{r}ava's theory admits  Lifshitz's  scale invariance: $\vec{r} \rightarrow b \vec{r}, \quad t \rightarrow b^{q} t$, and, after this, it is referred to as Ho\v{r}ava-Lifshitz (HL) gravity.  Actually, it has anistropic scaling in the UV domain, since it is $q=3$, while relativistic scaling with $q=1$ is recovered at large distances (IR).

Among the other issues, spherically symmetric solutions in HL gravity have been investigated in details\cite{spherical1,spherical2,spherical3,spherical4}, also in five dimensions\cite{spherical5d} (as for slowly rotating solutions, see Reg. \refcite{rot1,rot2}); in particular, Kehagias and Sfetsos\cite{KS} obtained an asymptotically flat and  static spherically symmetric solution that can be considered the analog of Schwarzschild solution in GTR.  An open issue in HL theory pertains the role of matter and its coupling to gravity, which has not been clarified in  full details. As a consequence, the motion of free particles in HL gravity is not trivial: since the fact that particles move along geodesics is not granted, it is possible to expect deviations from geodesic motion\cite{geo1,geo2,geo3,capasso}. However, many authors (see e.g. Ref. \refcite{light} and references therein)  assumed that free particles followed geodesics of KS  metric to  focus on many issues  HL gravity, such as gravitational lensing,  quasi-normal modes, accretion disks and so on (moreover, an accurate analysis of KS geodesics can be found in Ref. \refcite{geoKS}). In other words, they used KS solution as a toy model to study some fundamental aspects of the theory.

In this paper, starting from the same assumption, we aim at constraining  the dimensionsless  parameter $\psi_0$ of KS metric. In previous papers\cite{ijmpa,TOAJ}  we obtained  bounds  both from solar system and extra solar system observations of  orbital motions; other constraints were derived by studying light deflection\cite{light} and analyzing  the impact on  the classical GTR tests\cite{tiberiu}. Here we show that tighter results can be obtained by considering the perturbation $\Delta\rho$ induced by the KS solution on the two-body range $\rho$ for a pair of test particles A and B orbiting the same central body of mass $M$.

The plan of the work is as follows: in Section \ref{anale}, \textcolor{black}{after a brief introduction to the KS solution in the context of HL gravity,} we will analytically work out the perturbation $\Delta\rho$ induced by the KS solution on the two-body range $\rho$, which is a very accurate, direct and unambiguous observable in solar system planetary studies. We will try to make such a part as more self-consistent as possible, in view of a readership which may not be fully acquainted  with the subtleties of celestial mechanics applied to fundamental physics, an endeavor  requiring interdisciplinary knowledge across different fields. In this way the reader has the possibility of following autonomously the future developments of such kind of investigations, and to apply the present approach to other exotic long-range modified models of gravity as well. In Section \ref{osservazioni} we compare our theoretical predictions to the  range residuals $\delta\rho$ constructed with the latest planetary ephemerides applied to recent, accurate data sets of some planets of the solar system.
Section \ref{conclusioni} is devoted to summarizing our findings.

\section{Analytical computation of the two-body range perturbation}\label{anale}

\textcolor{black}{The starting point to obtain the KS solution is the action of the theory, given by\footnote{\textcolor{black}{Latin indices  run from 1 to 3, and refer to space components, while Greek run from 0 to 3, and refer to spacetime components.}}}
\textcolor{black}{
\begin{eqnarray}
S &= & \int dt d^3 x
\sqrt{g}N\left\{\frac{2}{\kappa^2}\left(K_{ij}K^{ij}-\lambda
K^2\right)-\frac{\kappa^2}{2w^4}C_{ij}C^{ij}+\frac{\kappa^2
\nu}{2w^2}\epsilon^{ijk} R^{(3)}_{i\ell}
\nabla_{j}R^{(3)\ell}{}_k \right.
\nonumber \\
&&\left. -\frac{\kappa^2\nu^2}{8} R^{(3)}_{ij} R^{(3)ij}+\frac{\kappa^2
\nu^2}{8(1-3\lambda)} \left[\frac{1-4\lambda}{4}\left(R^{(3)}\right)^2+\Lambda_W R^{(3)}-3
\Lambda_W^2\right]+\nu^4 R^{(3)}\right\}.\
\label{eq:action}
\end{eqnarray}}
\textcolor{black}{In \rfr{eq:action}, $g$ is the determinant of the metric tensor $g_{\gamma\delta}$, $K_{ij}$ and $C^{ij}$ are  the second fundamental form and the Cotton tensor, respectively, whose expressions are}
\textcolor{black}{\beq K_{ij}=\frac{1}{2N}\left(\dot{g}_{ij}-\nabla_i
N_j-\nabla_jN_i\right), \label{eq:Kij}
 \eeq}
\textcolor{black}{and}
\textcolor{black}{\beq
 C^{ij}=\epsilon^{ik\ell}\nabla_k
\left(R^{(3)j}{}_\ell-\frac{1}{4}R^{(3)} \delta^j_\ell\right), \label{eq:Cij}
\eeq}
\textcolor{black}{in terms of time derivative of the metric of the three-dimensional spatial slices $g_{ij}$, the lapse $N$ and the shift $N^i$, according to the standard ADM formalism (see e.g. Ref. \refcite{MTW}, Ch. 21), while $R^{(3)}_{ij}$ and $R^{(3)}$ are the Ricci tensor and scalar of the three-geometry. The parameters $\kappa,\lambda,w$ are coupling constants,
while $\Lambda_W^{2}$ is proportional to the cosmological constant, and\footnote{\textcolor{black}{Notice that Kehagias and Sfetsos used the symbol $\mu$ instead of $\nu$; in the present paper $\mu$ denotes the gravitational parameter of the central body which is the product of the Newtonian gravitational constant by its mass. See below, after \rfr{eq:metricas}.}} the constant $\nu$ has been introduced to modify the original HL action (see below). It is interesting to point out that standard GTR is recovered if the  running parameter $\lambda$ has a particular value, i.e. if $\lambda=1$ is an IR fixed point.}\textcolor{black}{\footnote{\textcolor{black}{However, it is important to stress that  even though  for $\lambda=1$ the action reduces to the one of GTR,  the theory possesses an extra degree of freedom which becomes strongly coupled when $\lambda$ goes to 1 (see e.g. Ref. \refcite{blas} and references therein).}}}
\textcolor{black}{Since the natural IR vacuum of the original HL theory is anti-de Sitter, in order to investigate the existence of limits of the theory with a Minkowski vacuum, Kehagias and Sfetsos deformed the theory with  the term $\nu^4 R^{(3)}$ and, furthermore, they considered the limit $\Lambda_{W} \to 0$.  In this limit, and looking for a statically symmetric solution in the form}
\textcolor{black}{
\beq
(ds)^2=N(r)^2(cdt)^2-\frac{(dr)^2}{f(r)}-r^2 \left(d\Phi\right)^{2},   \label{eq:metrica00}
\eeq}
\textcolor{black}{with $(d\Phi)^{2}\doteq (d\theta)^{2}+\sin^{2} \theta (d\varphi)^{2}$,  from (\ref{eq:action}) it is possible to obtain the corresponding Lagrangian:}
\textcolor{black}{\begin{equation}
{\cal{L}}=\frac{\kappa^2\nu^2}{8(1-3\lambda)}\frac{N}{\sqrt{f}}\left[(2\lambda-1)\frac{(f-1)^2}{r^2}
-2\lambda\frac{f-1}{r}f'+\frac{\lambda-1}{2}f'^2-2 \psi
(1-f-rf')\right], \label{eq:lagrangian0}
\end{equation}}
\textcolor{black}{where $\psi=8 \nu^2(3\lambda-1)/\kappa^2$. As for the $\lambda=1$ case, we get that  $\psi=16\nu^2/\kappa^2$. By solving the equations of motion, the asymptotically flat KS solution turns out to be} \textcolor{black}{
\beq (ds)^{2}= e^{\alpha(r)}(c dt)^{2} - e^{\beta(r)}(d r)^{2} -r^2(d\Phi)^{2}, \label{eq:metrica01} \eeq}
\textcolor{black}{where} \textcolor{black}{
\beq e^{\alpha(r)}=e^{-\beta(r)}=1+\psi r^{2}-\sqrt{\psi^{2}r^{4}+4 \psi \frac{\mathcal{\mu}}{c^{2}} r}, \label{eq:metricas} \eeq} \textcolor{black}{and $\mu\doteq GM$ is the \textcolor{black}{product of the Newtonian constant of gravitation $G$ times the mass $M$ of the central body which acts as source of the gravitational field}.\footnote{\textcolor{black}{Notice that Kehagias and Sfetsos used the symbol $\omega$ instead of $\psi$; we use the latter since the symbol $\omega$ might be confused with the pericenter of the orbit of a test particle, which is one of the standard Keplerian orbital elements: see the discussion after \rfr{nu}.}}}

\textcolor{black}{As we showed in Ref. \refcite{ijmpa}, in typical Solar System conditions, the metric (\ref{eq:metricas}) can be written in the form ``Schwarzschild plus corrections''. To this end, we set $\psi \doteq \frac{c^{4}\psi_0}{\mathcal{\mu}^2}$, where $\psi_{0}$ is a dimensionless parameter; then, we can see that in the limit}
\textcolor{black}{\beq \psi_0\gg 4\left(\frac{\mathcal{\mu}}{c^{2}r}\right)^3\approx 4\times 10^{-24}\label{condiz}\eeq} \textcolor{black}{for $M=M_{\odot}$ and $r\approx 1$ AU, we obtain}
\textcolor{black}{\beq e^{\alpha(r)}\approx 1-\frac{2\mathcal{\mu}}{c^{2}r} + \frac{2}{\psi_0}\left(\frac{\mathcal{\mu}}{c^{2}r}\right)^4.\eeq}

\textcolor{black}{As a consequence,  on introducing the  isotropic radial coordinate $\bar r$, the metric (\ref{eq:metricas}) becomes}
\textcolor{black}{\beq
(ds)^{2}=\left(1-\frac{2\mu }{c^{2} \bar r}+\frac{2\mu^{4}}{\psi_0 c^{8} \bar r^{4}} \right)(c dt)^{2}-\left(1+\frac{2\mu}{c^{2} \bar r}-\frac{1}{2}\frac{\mu^{4}}{\psi_0 c^{8} \bar r^{4}} \right) \left[(d \bar r)^{2}+ \bar r^{2}(d\Phi)^{2}\right]. \label{eq:metriciso}
\eeq}
\textcolor{black}{In the following, for the sake of simplicity, we will re-name $\bar r$ as $r$.}

\textcolor{black}{We point out that, even if here we are studying the KS solution obtained in the context of HL gravity, our results can be applied  as well to an arbitrary perturbation of the Schwarzschild solution in the form $e^{\alpha(r)}\approx 1-\frac{2{\mu}}{c^{2}r} + \frac{2\varepsilon}{r^{4}}$
where $\varepsilon$ is a suitable perturbation parameter. Also, for the sake of clarity, we stress that in performing our calculations we do assume that  particles move along geodesic, which is not, in general, granted in HL gravity, where also Lorentz invariance does not always hold. In summary, we use a relativistic post-Newtonian approach on the KS metric and the constraints that we obtain consistently apply to the metric parameters, rather than to the HL theory which, as we have seen, shows some inconsistencies.}

From (\ref{eq:metriciso})\textcolor{black}{, in addition to the usual Newtonian and general relativistic Schwarzschildian terms,} it is possible to obtain the corresponding KS acceleration, which reads (see also Ref. \refcite{ruggiero10})
\begin{equation}
\vec{A}_{\psi_0} = \frac{\mu^4}{\psi_0 c^6 r^5}\left[\left(4+\frac{v^2}{c^2}\right)\hat{\textcolor{black}{R}}-10\left(\frac{v_R}{c^2}\right)\vec{v}\right],\label{equazza}
\end{equation}
where $\hat{\textcolor{black}{R}}$ is the unit vector radially directed towards the test particle [see below, \rfr{erre}], while $v_R$ is the component of its velocity $\vec{v}$ along the radial direction (see below, \rfr{velaz}). \textcolor{black}{Notice that \rfr{equazza} is only valid at post-Newtonian order: it can be viewed as a small correction to the main Newtonian monopole term, to be treated with, e.g., the standard perturbative method by Gauss [see below, \rfr{Gauss}]. To this aim, let us notice that it usually refers to the Newtonian-Keplerian orbit assumed as unperturbed, reference trajectory. In principle, it is possible to assume as reference path a fully post-Newtonian one\cite{Calura1,Calura2}, and work out the effects of a given  small extra-acceleration like \rfr{equazza} with respect to it according to the perturbative scheme set up by the authors of Refs. \refcite{Calura1,Calura2}, which is a general relativistic generalization of another standard perturbative approach based on the planetary Lagrange equations\cite{Murr}. Actually, it is, in practice, useless since the only addition with respect to the orbital effects like, e.g., the precession of the pericenter, resulting from the standard scenario would consist of further, small  mixed GTR-perturbation orbital effects. \textcolor{black}{To be more explicit,  it can be shown that further acceleration terms of the order of\footnote{\textcolor{black}{They come from the term $-(c^2/2)h_{ik}h_{00,k},\ i=1,2,3$ in the post-Newtonian equations of motion\cite{Brum}.}} $\mu^5 c^{-8} r^{-6} \psi_0^{-1}$ and $\mu^8 c^{-14} r^{-9} \psi_0^{-2}$ would appear. The bounds on $\psi_0$ that we will obtain in Section \ref{osservazioni} will show, a posteriori, that they are, in fact, completely negligible in such a way that their inclusion would only make the calculation more cumbersome, without substantially affecting the constraints on $\psi_0$.}}

In order to perturbatively work out the orbital effects of \rfr{equazza}, it must, first,  be evaluated onto the unperturbed Keplerian ellipse\cite{Murr}
\eqi r=\rp{a(1-e^2)}{1+e\cos f},\lb{keple}\eqf where $a$ and $e$ are the semimajor axis and the eccentricity, respectively, characterizing its size and shape\footnote{It is  $0\leq e< 1$; circular orbits correspond to $e=0$.},  while $f$ is the true anomaly  reckoning the instantaneous position of the test particle  with respect to the point of closest approach to the central body, generally dubbed pericenter. To this aim, \rfr{equazza} has to be projected onto the radial ($R$), transverse ($T$) and normal ($N$) directions of an orthonormal frame co-moving with the orbiter\cite{Murr}. It is spanned by the unit vectors \textcolor{black}{$\{\hat{R},\hat{T},\hat{N}\}$} whose components with respect to a locally inertial frame $\{\hat{\i},\hat{\j},\hat{{\rm k}}\}$ centered in $M$ are\cite{cheng}
\begin{equation}
\hat{\textcolor{black}{R}}=\left\{
\begin{array}{lll}
\cos\Omega \cos u-\cos I\sin\Omega\sin u, \\ \\
\sin\Omega\cos u+\cos I\cos\Omega\sin u, \\ \\
\sin I\sin u,
\end{array}\lb{erre}
\right.
\end{equation}
\begin{equation}
\hat{\textcolor{black}{T}}=\left\{
\begin{array}{lll}
-\cos\Omega\sin u -\cos I\sin\Omega\cos u, \\ \\
-\sin\Omega\sin u+\cos I\cos\Omega\cos u, \\ \\
 \sin I\cos u,
\end{array}\lb{tau}
\right.
\end{equation}
\begin{equation}
\hat{\textcolor{black}{N}}=\left\{
\begin{array}{lll}
\sin I\sin\Omega, \\ \\
-\sin I\cos\Omega, \\ \\
 \cos I.
\end{array}\lb{nu}
\right.
\end{equation}
In \rfr{erre}-\rfr{nu} $I$ and $\Omega$ are the inclination of the orbital plane with respect to the reference $\{x,y\}$ plane and the longitude of the ascending node, respectively\textcolor{black}{: $\Omega$ is an angle in the reference $\{x,y\}$ plane delimited by the reference $x$ axis, usually taken coincident with the mean equinox at the epoch J2000.0 in planetary data analyses, and the line of the nodes, i.e. the intersection between the reference $\{x,y\}$ plane and the orbital plane. The angle} $u\doteq \omega +f$ is the argument of latitude in which the angle $\omega$ is the argument of pericenter:\textcolor{black}{ $\ \omega$  is an angle in the orbital plane reckoning the location of the point of closest approach to $M$ with respect to the line of the nodes.} Basically, $I,\Omega,\omega$ can be regarded as the three Euler angles determining the orientation of a rigid body, i.e. the unperturbed orbit\footnote{In the Keplerian two-body problem the ellipse neither changes its shape nor its size\textcolor{black}{, i.e., the semimajor axis $a$ and the eccentricity $e$ remain constant. M}oreover, it keeps its spatial orientation unchanged\textcolor{black}{, i.e., $I,\Omega,\omega$ do not change either}.}, in the inertial space\cite{Murr}.

By recalling that the $R-T-N$ components of the unperturbed Keplerian test particle's velocity are\cite{Murr}
 \begin{equation}
\left\{
\begin{array}{lll}
v_R & = & \frac{nae\sin f}{\sqrt{1-e^2}}, \\ \\
v_T & = & \frac{na \left(1+e\cos f\right)}{\sqrt{1-e^2}}, \\ \\
v_N & = & 0,
\end{array}\lb{velaz}
\right.
\end{equation}
where $n\doteq \sqrt{\mu/a^3}$ is the unperturbed Keplerian mean motion,
 it turns out that the $R-T-N$ components of \rfr{equazza} are
\begin{equation}
\left\{
\begin{array}{lll}
A_R & = & \frac{\mu^4 \left(1+e\cos f\right)^5}{\psi_0 c^6 a^5 (1-e^2)^6} \left\{ 4\left(1-e^2\right)+\frac{\mu}{c^2 a}\left[\left(1-4e^2\right) +e\left(2\cos f +5 e \cos 2 f \right)\right]   \right\}, \\ \\
A_T & = & -\frac{10 e\mu^5(1+e\cos f)^6 \sin f}{\psi_0 c^8 a^6  (1-e^2)^6}, \\ \\
A_N & = & 0.
\end{array}\lb{RTNacc}
\right.
\end{equation}
\textcolor{black}{Notice that, in the limit $e\rightarrow 0$, it turns out that, contrary to $A_R$,  $A_T\rightarrow 0$: moreover, $v_R\rightarrow 0$, while $v_T$ does not vanish. Actually, there is nothing strange or pathological in that: just as in the\footnote{\textcolor{black}{It also holds in GTR since, as it is well known, the Schwarzschild acceleration becomes radial for a circular orbit.}} Newtonian case,  the acceleration experienced by the test particle is entirely radial, and its velocity vector is tangential to the circular trajectory, changing only its direction.}
An inspection of the content within the curly brackets of $A_R$ in \rfr{RTNacc} shows that the second term  is, usually, several orders of magnitude smaller that the first one, which is of the order of unity\textcolor{black}{: indeed, in the case of the Sun and, say, Mercury it is
\eqi \rp{\mu_{\odot}}{c^2 a_{\mercury}} = 2\times 10^{-8}.\eqf} Moreover, in the limit $e\rightarrow 0$, valid   for circular orbits, \textcolor{black}{ as we already noticed,} $A_T$ vanishes, while $A_R$ reduces to a constant term depending on the test particle's orbital radius as $a^{-5}$ to the leading order. 

At this point, the perturbations of all the six Keplerian orbital elements\cite{Murr}, including also the mean anomaly $\mathcal{M}$, have to be computed by means of the standard Gauss perturbative equations\cite{Murr}
\begin{equation}
\left\{
\begin{array}{lll}
\dert a t & = & \rp{2}{n\sqrt{1-e^2}} \left[e A_R\sin f +A_{T}\left(\rp{p}{r}\right)\right],\\   \\
\dert e t  & = & \rp{\sqrt{1-e^2}}{na}\left\{A_R\sin f + A_{T}\left[\cos f + \rp{1}{e}\left(1 - \rp{r}{a}\right)\right]\right\},\\  \\
\dert I t & = & \rp{1}{na\sqrt{1-e^2}}A_N\left(\rp{r}{a}\right)\cos u,\\   \\
\dert\Omega t & = & \rp{1}{na\sin I\sqrt{1-e^2}}A_N\left(\rp{r}{a}\right)\sin u,\\    \\
\dert\omega t & = &\rp{\sqrt{1-e^2}}{nae}\left[-A_R\cos f + A_{T}\left(1+\rp{r}{p}\right)\sin f\right]-\cos I\dert\Omega t,\\   \\
\dert {\mathcal{M}} t & = & n - \rp{2}{na} A_R\left(\rp{r}{a}\right) -\sqrt{1-e^2}\left(\dert\omega t + \cos I \dert\Omega t\right).
\end{array}\lb{Gauss}
\right.
\end{equation}
 In \rfr{Gauss} $p\doteq a(1-e^2)$ is the semi-latus rectum.
It turns out that by  inserting \rfr{RTNacc} in \rfr{Gauss} and integrating it by means of\cite{Murr}
\eqi dt = \rp{1}{n}\left(\rp{r}{a}\right)^2 \rp{1}{\sqrt{1-e^2}} df\eqf from an initial value $f_0$ to a generic $f$ allows to obtain exact expressions for the shifts $\Delta a, \Delta e, \Delta I, \Delta \Omega, \Delta\omega, \Delta \mathcal{M}$. Since they are extremely cumbersome, we will not explicitly display them. The inclination $I$ and the node $\Omega$ are not affected since, according to \rfr{RTNacc}, $A_N=0$.

The further step consists of putting such formulas  in the following expressions for the $R-T-N$ perturbations of the position vector $\vec{r}$ taken from Ref.~\refcite{Caso}
\begin{equation}
\left\{\begin{array}{lll}
\Delta R &=&\left(\rp{r}{a}\right)\Delta a - a \cos f\Delta e +\rp{ae\sin f}{\sqrt{1-e^2}}\Delta{\mathcal{M}}, \\ \\
\Delta T &=& a\sin f\left[1+\rp{r}{a(1-e^2)}\right]\Delta e + r(\cos I\Delta\Omega + \Delta\omega)+\left(\rp{a^2}{r}\right)\sqrt{1-e^2}\Delta{\mathcal{M}},\\ \\
\Delta N &=& r(\sin u\Delta I-\cos u\sin I\Delta\Omega).
\end{array}\lb{dnorm}
\right.
\end{equation}
In the unperturbed case, there is only the radial component, i.e. $\vec{r} = r\hat{\textcolor{black}{R}}$, with $r$ and $\hat{\textcolor{black}{R}}$ given by \rfr{keple} and \rfr{erre}, respectively.
In the case of \rfr{RTNacc},
only the radial and transverse shifts occur because $\Delta N$ is induced by  $\Delta I$ and $\Delta\Omega$ which, as already noted, vanish. Also in this case, we do not display the resulting exact formulas because of their excessive length and complexity.

Finally, we are ready to analytically work out the perturbation $\Delta\rho$ of the two-body range  $\rho$ between two test particles A and B orbiting the same central body. Indeed, the range is a direct and accurate observable which is widely used in astronomical studies either by directly ranging from A $=$ Earth to the surface of the target body  B or to a spacecraft orbiting it.
From\cite{cheng}
\eqi
\left\{
\begin{array}{lll}
\rho^2 &=&\left(\vec r_{\rm A}-\vec r_{\rm B}\right)\cdot\left(\vec r_{\rm A}-\vec r_{\rm B}\right),\\ \\
\hat{\rho} &\doteq & \rp{\left(\vec r_{\rm A}-\vec r_{\rm B}\right)}{\rho},
\end{array}
\right.
\eqf
to be evaluated onto the unperturbed Keplerian ellipses
of the two test particles A and B,
it follows that, for a generic perturbation, the range shift $\Delta \rho$ is\cite{cheng}
\eqi\Delta\rho=\left(\Delta\vec r_{\rm A}-\Delta\vec r_{\rm B}\right)\cdot\hat{\rho},\lb{range}\eqf
where
\eqi\Delta \vec{r_j}=\Delta R_j\ \hat{
\textcolor{black}{R_j}
}+\Delta T_j\ \hat{\textcolor{black}{T_j}}+\Delta N_j\ \hat{\textcolor{black}{N_j}},\ j={\rm A,B}.\eqf
Note that, in general, all the components of the perturbations of the position vectors of both A and B are needed since, in general, $\hat{\textcolor{black}{R_i}} \cdot\hat{\textcolor{black}{T_j}}$ and $\hat{\textcolor{black}{R_i}} \cdot\hat{\textcolor{black}{N_j}}$ do not vanish for $i,j={\rm A,\ B},\ i\neq j$. In other words, $\Delta\rho$ is not simply the difference between $\Delta R_{\rm A}$ and $\Delta R_{\rm B}$, as it might  intuitively be guessed.

In order to conveniently plot \rfr{range} as a function of time, we will use the following useful approximate equation for the true anomalies of both A and B\cite{Murr}
\eqi
\begin{array}{lll}
f&\simeq &{\mathcal{M}} + 2e\sin {\mathcal{M}} +\rp{5}{4}e^2\sin 2{\mathcal{M}}
+ \\ \\
& + & e^3\left(\rp{13}{12}\sin 3{\mathcal{M}} -\rp{1}{4}\sin {\mathcal{M}}\right)+e^4\left(\rp{103}{96}\sin 4{\mathcal{M}} - \rp{11}{24}\sin 2{\mathcal{M}}\right);
\end{array}\lb{anom}
\eqf
indeed, ${\mathcal{M}}\doteq n(t-t_p),$ where $t_p$ is the time of the passage at pericenter.
%
\section{Confrontation with the observations}\lb{osservazioni}
Our analytical calculations, suitably compared with latest observational results\cite{Fienga10}, allow to put tighter constraints on the KS parameter $\psi_0^{\odot}$ with respect to those inferred in Ref.~\refcite{ijmpa} from the corrections\footnote{Here $\varpi\doteq\Omega+\omega$ denotes the longitude of pericenter\textcolor{black}{: it is a \virg{dogleg} angle since, in general, it does not lie in a plane.}.} $\Delta\dot\varpi$ to the standard secular perihelion precessions.

By preliminarily working  in the limit  $e\rightarrow 0$, a naive order-of-magnitude evaluation can be obtained by interpreting $A_R$  in \rfr{RTNacc} as due to an \virg{effective} gravitational parameter
\eqi\overline{\mu}(a)\doteq\mu+\rp{4\mu^4}{\psi_0 c^6 a^3}\eqf
in the Newtonian monopole. Since $\Delta \mu\doteq \overline{\mu}(a)-\mu$ cannot be larger than the  accuracy $\sigma_{\mu}$ with which the  gravitational parameter of the central body of the system under consideration is known, it turns out
\eqi\psi_0\gtrsim \rp{4\mu^4}{\sigma_{\mu} c^6 a^3}.\lb{grumi}\eqf Note that \rfr{grumi} tells us that the tightest constraints come from  test particles orbiting very close to  highly massive central bodies. A step further consists of taking the average of $A_{\psi_0}$ over an orbital period: in doing that we may neglect the terms of order $\mathcal{O}(c^{-8})$ in \rfr{RTNacc}, but we do not set $e=0$. The result, exact to $\mathcal{O}(c^{-6})$, is
\eqi\overline{\mu}(a,e)\doteq \mu+\rp{2(2+3 e^2)\mu^4}{\psi_0 c^6 a^3(1-e^2)^{7/2}}.\eqf Thus,
\eqi \psi_0\gtrsim \rp{2(2+3 e^2)\mu^4}{\sigma_{\mu} c^6 a^3(1-e^2)^{7/2}},\lb{grumi2}\eqf according to which a high eccentricity contributes to further strengthen the constraints on $\psi_0$.

By applying \rfr{grumi2} to the eight major planets of the solar system we have the bounds summarized in Table \ref{tavola}.
\begin{table}[ph]
\tbl{Lower bounds of the KS parameter $\psi_0^{\odot}$ from \rfr{grumi2} with $\sigma_{\mu}=1\times 10^9$ m$^3$ s$^{-2}$,  from Table 3 of Ref.~\protect\refcite{Fienga10}.\label{tavola}
}
{\begin{tabular}{@{}cc@{}} \toprule
Planet & $\psi_0^{\odot}$\\
 \colrule
Mercury & $1.1\times 10^{-11}$\\
Venus & $1.3\times 10^{-12}$\\
Earth & $5.1\times 10^{-13}$\\
Mars & $1.4\times 10^{-13}$\\
Jupiter & $3.6\times 10^{-15}$\\
Saturn & $5.8\times 10^{-16}$\\
Uranus & $7.2\times 10^{-17}$\\
Neptune & $1.8\times 10^{-17}$\\
%
\botrule
\end{tabular}}
\end{table}
They are tighter than those in Table 3-Table 4 of Ref.~\refcite{ijmpa} by $1-6$ orders of magnitude, especially for the outer planets.
As expected, the tightest constrain comes from Mercury, with \eqi\psi_0^{\odot}\geq 1.1\times 10^{-11},\eqf which is about $1.4$ times larger than the bound in Table 3 of Ref.~\refcite{ijmpa}. The use of the less accurate \rfr{grumi} would have yielded the weaker constraint
\eqi\psi_0^{\odot}\geq 8.8\times 10^{-12}.\eqf

However, such a back-to-the-envelope approach cannot substitute a more rigorous treatment. Thus, in the following we will compare our predicted results for the KS range perturbation $\Delta \rho$ to the post-fit range residuals of some planets produced\footnote{\textcolor{black}{It is just the case to recall that GTR was fully modelled, so that such residuals account, in principle, for any other un-modelled dynamical effects like  the KS perturbations.}} with the latest solar system ephemerides\cite{Fienga10}. As we will see, even tighter constraints will result from such a detailed analysis.

It is interesting to note that if we apply \rfr{grumi2} to the system\cite{BH} constituted by the S2 star ($a=1031.69$ au, $e=0.8831$) orbiting the Supermassive Black Hole ($\mu_{\bullet}=4\times 10^6\mu_{\odot},\ \sigma_{\mu_{\bullet}}=0.5\times 10^6\mu_{\odot}$) located at the center of the Galaxy we get
\eqi\psi_0^{\bullet}\geq 8\times 10^{-10}.\eqf The bound from \rfr{grumi} would be
 \eqi\psi_0^{\bullet}\geq 2\times 10^{-12}, \eqf i.e. 446 times smaller; this clearly shows how \rfr{grumi} is inadequate for highly eccentric orbits.
 In order to correctly interpret such result it should be kept in mind that the KS parameter $\psi_0$, contrary to, say, $G$ and $c$, is not a universal constant of Nature; in general, it may vary from a space-time to another being, thus, connected with the body which acts as source of the gravitational field.

 Moving to the planetary range residuals, the authors of Ref.~\refcite{Fienga10} processed some radiometric data from the Messenger spacecraft during its recent flybys of Mercury in 2008-2009 with the INPOP10a ephemerides; according to Table 1 of Ref.~\refcite{Fienga10}, the resulting mean and standard deviation of the produced range residuals are
 \eqi \delta\rho = -0.6\pm 1.9\ {\rm m}.\eqf
 In order to constrain $\psi_0^{\odot}$ from such data we will use the analytical results of Section \ref{anale} to calculate a time series for $\Delta\rho$ with the same numerical characteristics. The result is depicted in Figure \ref{messenger}:
 \begin{figure*}[ht!]
\centering
\begin{tabular}{c}
\epsfig{file=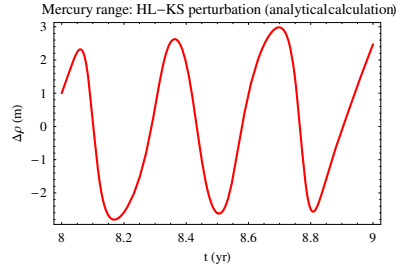,width=0.45\linewidth,clip=}
\end{tabular}
\caption{Analytically computed Earth-Mercury  range perturbation $\Delta\rho$ caused by KS for $\psi_0^{\odot} = 7.2\times 10^{-10}$ over $\Delta t=1$ yr. The initial conditions, corresponding to J2000.0 (mean ecliptic and equinox at such an epoch), have been retrieved from the NASA-JPL WEB interface HORIZONS. }\lb{messenger}
\end{figure*}
 it has been obtained for
 \eqi\psi_0^{\odot}= 7.2\times 10^{-10},\lb{gonza}\eqf and has just
 \eqi\Delta\rho =  0.1\pm 1.9\ {\rm m}.\lb{joppa}\eqf
 We checked our analytical result by simultaneously integrating the equations of motion of the Earth and Mercury with and without \rfr{equazza} for the same initial conditions used for producing Figure \ref{messenger}; it turns out that the resulting numerically computed KS range signal shows an excellent agreement with the analytical one, both qualitatively and quantitatively. Thus, we can  reasonably be confident in our analytical approach.
 A  value for $\psi_0^{\odot}$ smaller than that in \rfr{gonza} would generate a signal  with a larger standard deviation in \rfr{joppa}. Thus, we take the result of \rfr{gonza} as our $1-\sigma$ lower bound on $\psi_0^{\odot}$. It is about two orders of magnitude tighter than the constraint in Table 3 of Ref.~\refcite{ijmpa}. It is also 65 times larger than the naive result of Table \ref{tavola}. However, we remark that HL gravity was not modelled at all in the INPOP10a ephemerides, which, as usual, only took into account the main well established, standard Newtonian and Einsteinian effects. Thus, if HL really existed in Nature, its signature might have been partially removed by the signal in the estimation of the state vectors during the data reduction process. In other words, there is, in principle, the possibility that $\psi_0^{\odot}$ may actually be smaller than \rfr{gonza}. As for any other putative exotic models of gravity, it would be necessary to explicitly include HL  in the force models used to process the data and repeat the entire fitting procedure with such a modified dynamical theory: however, such a task is definitely not a trivial one and would require a dedicated analysis by professional teams of astronomers. It is outside the scopes of the present paper. On the contrary, the approach outlined here is relatively easy to implement, and can be extended to other modified models of gravity as well.

 A similar analysis can be done for Mars. Indeed, its range residuals, built with the ephemerides INPOP10a and the data of the Mars Global Surveyor (MGS) spacecraft over a time span 10 yr long (1998-2008), yield\cite{Fienga10}
 \eqi \delta \rho = 0.5\pm 1.9\ {\rm m}.\eqf
 Figure \ref{mgs} displays the analytically computed KS range signal over the same time interval for \eqi\psi_0^{\odot} = 9\times 10^{-12}.\lb{uiop}\eqf
 \begin{figure*}[ht!]
\centering
\begin{tabular}{c}
\epsfig{file=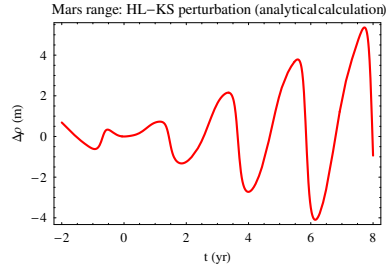,width=0.45\linewidth,clip=}
\end{tabular}
\caption{Analytically computed Earth-Mars  range perturbation $\Delta\rho$ caused by KS for $\psi_0^{\odot} = 9\times 10^{-12}$ over $\Delta t=10$ yr. The initial conditions, corresponding to J2000.0 (mean ecliptic and equinox at such an epoch), have been retrieved from the NASA-JPL WEB interface HORIZONS. }\lb{mgs}
\end{figure*}
It turns out that
\eqi \Delta\rho = 0.3\pm 2\ {\rm m}.\eqf
A numerical integration confirms such features, both qualitatively and quantitatively.
The constraint of   \rfr{uiop} is 64 times larger than the approximate one in Table \ref{tavola}. With respect to Table 3 of Ref.~\refcite{Fienga10}, \rfr{uiop} is more stringent by three orders of magnitude.

 Another planet  whose orbit's knowledge has greatly been improved in recent years thanks to the continuous radiometric ranging to the Cassini spacecraft is Saturn.
 The range residuals over 3 years (2004-2007) produced with the INPOP10a ephemerides\cite{Fienga10} yield
 \eqi \delta\rho = 0.0\pm 17\ {\rm m}.\eqf  Figure \ref{cassini} depicts the analytically computed KS range perturbation for
 \eqi\psi_0^{\odot}= 1.7\times 10^{-12}:\lb{sbiro}\eqf
 \begin{figure*}[ht!]
\centering
\begin{tabular}{c}
\epsfig{file=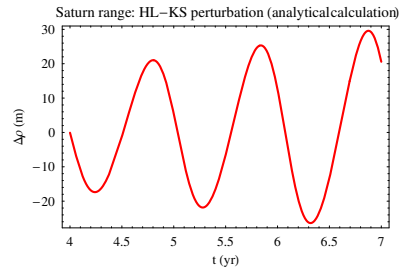,width=0.45\linewidth,clip=}
\end{tabular}
\caption{Analytically computed Earth-Saturn  range perturbation $\Delta\rho$ caused by KS for $\psi_0^{\odot} = 1.7\times 10^{-12}$ over $\Delta t=3$ yr. The initial conditions, corresponding to J2000.0 (mean ecliptic and equinox at such an epoch), have been retrieved from the NASA-JPL WEB interface HORIZONS. }\lb{cassini}
\end{figure*}
 it has
 \eqi\Delta \rho = 0.7\pm 17\ {\rm m}.\eqf Also in this case, a numerical integration fully confirms the characteristics of the analytical signal of Figure \ref{cassini}.
 The constraint of \rfr{sbiro} is about four orders of magnitude tighter than that in Table \ref{tavola}, and six orders of magnitude better than Table 4
 in Ref.~\refcite{ijmpa}: cfr. with the much smaller deviations in the case of Mercury and Mars amounting to about one order of magnitude.  It is likely so because the radiometric data from the inner planets played a major role in the estimation of the Sun's gravitational parameter\cite{Fienga10}.
 \textcolor{black}{As anticipated in Section \ref{anale}, the lower bounds derived for $\psi_0$ make negligible the mixed terms that we did not included in our calculation.}

Finally, we note that the solar system constraints on $\psi_0^{\odot}$ obtained in the present paper are more stringent than those derived by the authors of Ref.~\refcite{light} through the light deflection,  and by the authors of Ref.~\refcite{tiberiu} from the orbital motion of Mercury, the deflection
of light by the Sun, and the radar echo delays.
\section{Summary and conclusions}\lb{conclusioni}
In view of the fact that ranging from the Earth to some major bodies of the solar system is one of the most direct and accurate way to determine their orbits, in this paper we analytically worked out the effects that the Kehagias-Sfetsos solution of the Ho\v{r}ava-Lifshitz modified gravity at long distances have on the two-body range $\rho$. We successfully checked our analytical results by simultaneously integrating the equations of motion of the bodies A and B considered by including such exotic dynamical effects as well.

By comparing our predictions with the range-residuals $\delta\rho$ of Mercury, Mars and Saturn produced with the recent ephemerides INPOP10a we have been able to effectively constrain the dimensionless free parameter $\psi_0$ entering the Kehagias-Sfetsos equations in the case of the Sun. We obtained
$\psi^{\odot}_0\geq 7.2\times 10^{-10}\ ({\rm Mercury}),\ \psi^{\odot}_0\geq 9\times 10^{-12}\ ({\rm Mars}),\ \psi^{\odot}_0\geq 1.7\times 10^{-12}\ ({\rm Saturn})$. Such constraints are orders of magnitude better than those existing in literature.

In principle, one should re-process the entire planetary data set by explicitly modeling the Kehagias-Sfetsos dynamical effects in addition to the standard Newtonian-Einsteinian  ones and solving for a dedicated parameter in order to avoid the possibility that they, if unmodeled,  may be removed from the signal in the estimation of the initial conditions. However, such a task is very time-consuming and can effectively be implemented only by skilful and expert teams of specialists in astronomical data processing. On the other hand, the  relatively simpler approach outlined here can be easily and quickly reproduced, and extended to other exotic long-range modified models of gravity as well in order to yield reasonable evaluations of the magnitude of the effects of interest.
\section*{Acknowledgements}
One of us (L.I.) thanks D. Grumiller for useful hints about the use of the effective gravitational parameter. \textcolor{black}{Both the authors gratefully acknowledge the referee for her/his useful comments and critical remarks which helped in improving the manuscript.}


\begin{thebibliography}{00}

\bibitem{horava1} P. Ho\v{r}ava, \textit{Phys. Rev. D } \textbf{79}, 84008 (2009).

\bibitem{horava2} P. Ho\v{r}ava, \textit{Phys. Rev. Lett. } \textbf{102}, 161301 (2009).

\bibitem{sotiriou} T.P. Sotiriou, based on talk given at the 14th Conference on Recent Developments in Gravity (NEBXIV), Ioannina, Greece, 8-11 June 2010, arXiv:1010.3218.


\bibitem{spherical1} J. Alexandre and P. Pasipoularides, arXiv:1010.3634.

\bibitem{spherical2} D. Capasso, \textit{Phys. Rev. D} \textbf{82},124058 (2010).


\bibitem{spherical3} J. Greenwald, V. H. Satheeshkumar, A. Wang, \textit{JCAP} 1012:007 (2010).

\bibitem{spherical4} G. Koutsoumbas, P. Pasipoularides, \textit{Phys. Rev. D} \textbf{82}, 044046 (2010)

\bibitem{spherical5d} G. Koutsoumbas, E. Papantonopoulos, P. Pasipoularides, M. Tsoukalas,  \textit{Phys. Rev. D} \textbf{81}, 124014 (2010)

\bibitem{rot1} A. Ghodsi, E. Hatefi, \textit{Phys. Rev. D} \textbf{81}, 044016 (2010).

\bibitem{rot2} A. N. Aliev, \c{C}.  \c{S}ent\"{u}rk \textit{Phys. Rev. D} \textbf{82}, 104016 (2010).

	
\bibitem{KS}
A. Kehagias and K. Sfetsos, {\it Phys. Lett. B} {\bf 678}, 123 (2009).


\bibitem{geo1} S.K. Rama, arXiv:0910.0411.

\bibitem{geo2} A.E. Mosaffa, arXiv:1001.0490.

\bibitem{geo3} M. Eune and W. Kim, \textit{Mod. Phys. Lett. A} \textbf{25}, 2923 (2010).

\bibitem{capasso} D. Capasso and A.P. Polychronakos,  \textit{J. High En. Phys.} \textbf{02}, 068 (2010).

\bibitem{light}
M. Liu, J. Lu, B. Yu and J. Lu, \textit{Gen. Relativ. Gravit.} doi:10.1007/s10714-010-1123-0 (2010).

\bibitem{geoKS} B. Gwak, B.H. Lee, \textit{JCAP} 1009:031  (2010).

\bibitem{ijmpa}
L. Iorio and M. L. Ruggiero, \textit{Int. J. Mod. Phys. A.} \textbf{25}, 5399 (2010).

\bibitem{TOAJ}
L. Iorio and M. L. Ruggiero, \textit{Open Astron. J.} \textbf{3}, 167 (2010).

\bibitem{tiberiu}
T. Harko, Z. Kov\'{a}cs and F. S. N. Lobo, \textit{Proc. Roy. Soc. A}  \textbf{467}, 1390 (2011).

\bibitem{MTW} C.W. Misner, K.S. Thorne, J.A. Wheeler, \textit{Gravitation} (Freeman, San Francisco, 1973).

\bibitem{blas} D. Blas, O. Pujolas, S. Sibiryakov, \textit{JHEP} 0910:029 (2009).

\bibitem{ruggiero10} M. L. Ruggiero, arXiv:1010.2114.

\bibitem{Calura1}
Calura M., Fortini P., Montanari E., \textit{Phys. Rev. D} \textbf{56}, 4782 (1997).
%
\bibitem{Calura2}
Calura M., Fortini P., Montanari E., \textit{Class. Quantum Gravit.} \textbf{15}, 3121 (1998).


\bibitem{Brum}
V. A. Brumberg, Essential Relativistic Celestial Mechanics (Adam Hilger, Bristol, 1991).


\bibitem{Murr}
C. D. Murray and S. F. Dermott, \textit{Solar System Dynamics} (Cambridge University Press, Cambridge, 1999).

\bibitem{cheng}
M. K. Cheng,
\textit{J. of Geodesy} \textbf{76}, 169
(2002).

\bibitem{Caso}
S. Casotto,
\textit{Celest. Mech. and Dyn. Astron.}  \textbf{55},  209
(1993).


\bibitem{Fienga10}
A. Fienga, H.  Manche, P. Kuchynka, J. Laskar and M. Gastineau, arXiv:1011.4419.

\bibitem{BH}
S. Gillessen, F. Eisenhauer, T. K. Fritz, H. Bartko, K. Dodds-Eden, O. Pfuhl, T. Ott and R. Genzel,  \textit{Astrophys. J.} \textbf{707}, L114 (2009).






\end{thebibliography}
\end{document}